\documentclass[aps,prd,preprint]{revtex4-1}
\usepackage{graphicx}
\def\be{\begin{equation}} 
\def\ee{\end{equation}  }
\def\bea{\begin{eqnarray}}
\def\eea{\end{eqnarray}  }

\begin{document}
\title{White Holes in Einstein-Aether Theory}
\author{Ratindranath Akhoury}
\email{akhoury@umich.edu}
\affiliation{Michigan Center for Theoretical Physics, Randall Laboratory of Physics, University of Michigan, Ann Arbor, MI 48109-1120, USA}
\author{David Garfinkle}
\email{garfinkl@oakland.edu}
\affiliation{Dept. of Physics, Oakland University, Rochester, MI 48309, USA}
\affiliation{Michigan Center for Theoretical Physics, Randall Laboratory of Physics, University of Michigan, Ann Arbor, MI 48109-1120, USA}
\author{Nishant Gupta}
\email{nishash@umich.edu}
\affiliation{Dept. of Physics, University of Michigan, Ann Arbor, MI 48109-1120, USA}

\date{\today}

\begin{abstract}

We perform numerical simulations of gravitational collapse in Einstein-aether theory.  We find that under certain conditions, the collapse results in the temporary formation of a white hole horizon.
\end{abstract}

\maketitle

\section{Introduction}

Over the past several years there has been interest in gravitational theories with a dynamical vector field, including Einstein-aether theory\cite{jacobsonmattingly,deserfest}, TeVes\cite{teves}, and Ho{\v r}ava gravity\cite{horava} (which in certain cases can be regarded as a limiting case of Einstein-aether theory\cite{tedhorava}).  There are various motivations for these studies, including efforts to quantize gravity, or to provide an alternative explanation for those effects usually attributed to dark matter, or simply to provide an alternative to general relativity that could possibly be distinguished by observations of gravitational radiation.  

One interesting aspect of gravity is black holes, so it is natural to ask what are the properties of black holes in these theories.  Here one can find black holes by assuming a static spacetime with an event horizon and finding the corresponding solution of the field equations.  Or one can numerically simulate gravitational collapse and find the endstate of the process.  For Einstein-aether theory, both of these approaches have been used.\cite{aetherbh1,dgandted,barated} The approach of \cite{aetherbh1} was to assume a static, spherically symmetric spacetime with an event horizon.  These assumptions reduced the field equations of the theory to a set of coupled ordinary differential equations (for the fields as a function of the radial coordinate) which satisfied an appropriate set of boundary conditions (smoothness at the horizon and asymptotic flatness at infinity).  These equations could not be solved in closed form, but could be solved numerically.  Einstein-aether theory contains four free parameters, ${c_1}, {c_2}, {c_3}$ and $c_4$ representing coefficients of different terms in the action.  In\cite{aetherbh1} black hole solutions were found for certain ranges of the parameters; however, it was found that when $c_1$ was sufficiently large the method did not find black hole solutions. 

The fact that a black hole solution exists does not necessarily mean that the corresponding black hole is actually the endstate of gravitational collapse.  To explore this issue, in \cite{dgandted} simulations were performed of gravitational collapse in Einstein-aether theory in spherical symmetry with a scalar field providing the matter stress-energy.  Here it was found that with the $c_i$ in the range where \cite{aetherbh1} found black hole solutions, the endstate of gravitational collapse was indeed the corresponding black hole solution.  However, for the large values of $c_1$ for which the method of \cite{aetherbh1} failed to find black holes, the collapse simulations of \cite{dgandted} resulted in configurations in which the spatial derivatives of the fields became very large, which the authors of \cite{dgandted} speculated was a precursor of the formation of a naked singularity.  

This naked singularity interpretation was rendered somewhat dubious when the authors of \cite{barated} revisited the ordinary differential equation calculation of static black hole solutions with improved methods and succeeded in finding solutions for large values of $c_1$ where the previous treatment of \cite{aetherbh1} failed to find solutions.  Are these additional solutions the endstates of gravitational collapse? And if so, then why did the simulations of \cite{dgandted} fail to find them?  To answer these questions, we revisit the collapse simulation problem with our own improved numerical methods.  One possible reason for fields to have a large spatial derivative is simply an unfortunate choice of spatial coordinate.  We therefore perform for each gravitational collapse situation, two simulations: one with the radial coordinate used in \cite{dgandted} and one with a different radial coordinate.  In addition, we introduce a more general type of initial data and an improved outer boundary condition for better stability and to allow the simulation to run longer.  

Much to our surprise, we find that the collapse process for large $c_1$ produces neither black holes nor naked singularities.  Instead the collapse results in white holes! That is, during the collapse rather than a trapped surface forming, an anti-trapped surface forms instead.  This anti-trapped surface is a temporary phenomenon, and eventually the fields disperse.  Therefore the additional solutions of \cite{barated} are not the endstates of gravitational collapse.

In section II we present the relevant facts about the field equations of Einstein-aether theory, while section III treats the initial data used and the method by which the field equations are evolved.  (Detailed treatments of the equations of motion using the two different radial coordinates are given in appendicies).  In section IV we present our results.  Conclusions are given in section V.  

\section{Einstein-Aether Theory And Its Field Equations}
Einstein-aether theory \cite{jacobsonmattingly} is general relativity with a dynamical unit timelike vector field. This vector field cannot vanish; it picks out a preferred reference frame and thus the theory spontaneously breaks Lorentz invariance.  

The action  $S$ for Einstein-aether theory is chosen to be the most general, generally covariant functional of the spacetime metric $g_{ab}$ and the aether field $u^a$, involving no more than two derivatives. This results in four terms involving the aether field, and they are included with arbitrary coefficients.  The action takes the form\cite{jacobsonmattingly}:
\be
 S = \int \sqrt{- g}(L_{ae} + L_{matter}) \; d^{4}x, 
\ee
where, 
\be
 L_{ae} = {\frac {1}{16\pi G}}[{\cal R} - {K^{ab}}_{mn}{\nabla_a} {u^m}{\nabla_b}{u^n} + \lambda(g_{ab}u^{a}u^{b} + 1)]. 
\ee

Here $\cal R$ denotes the Ricci scalar, $G$ a parameter related to Newton's constant \cite{four}, $L_{matter}$ denotes the matter Lagrangian density and $\lambda$ is a Lagrange multiplier enforcing the condition that the aether field is unit timelike at all points of spacetime. ${K^{ab}}_{mn}$ is defined as:
\be
{K^{ab}}_{mn} = c_1g^{ab}g_{mn} + c_2\delta^a_m\delta^b_n + c_3\delta^a_n\delta^b_m - c_4u^au^bg_{mn},
\ee
where the $c_i$ are dimensionless constants. The Lorentzian signature used in this paper is ($-$,+,+,+) and the units are chosen so that the speed of light defined by the metric $g_{ab}$ is 1. Einstein-aether theory possesses spin-1, spin-2 and spin-0 massless modes that travel at speeds different from each other and from the speed of light, so that for a solution to describe a black hole in Einstein-aether theory, all of these wave modes must also be trapped in the region of the black hole.

We take the matter field to be a minimally coupled massless scalar field $\chi$, following the choice in \cite{dgandted}, with Lagrangian $-{\nabla _a}\chi{\nabla ^a}\chi$, which we scale by $\psi$ = $\chi \sqrt{16\pi G}$ to simplify the form of the field equations. The matter Lagrangian is then:

\be
 L_{matter} = {\frac {-1}{16\pi G}} \nabla_a\psi\nabla^a\psi. 
\ee

We choose the surfaces of constant time in our simulation to be those orthogonal to $u^a$. This choice is possible since we will be working in a spherically symmetric system for which $u^a$ is necessarily hypersurface orthogonal. 
As is usual in numerical relativity, the metric degrees of freedom are described in terms of the spatial metric, extrinsic curvature, lapse function and shift vector.  
The spatial metric $h_{ab}$ and extrinsic curvature $K_{ab}$ are given by
\bea
h_{ab} = g_{ab} + u_au_b, 
\label{hdef}
\\
K_{ab} = -{\frac 1 2}{{\cal L}_u} h_{ab},
\label{kdef}
\eea
where $\cal L$ denotes the Lie derivative.

Due to the property of spherical symmetry, the spacetime line element takes the form
\be
d {s^2} = - {\alpha ^2} d {t^2} + \gamma {{(dr + {\beta ^r} dt)}^2} + {R^2} (d{\theta ^2} + {\sin^2}\theta d {\phi^2})
\label{sphds2}
\ee
where $\alpha$ is the lapse function and $\beta ^a$ is the shift vector.  Comparison of eqns. (\ref{hdef}) and (\ref{sphds2}) shows that $\gamma ={h_{rr}}$ and ${R^2}={h_{\theta \theta}}={h_{\phi \phi}}/{\sin^2}\theta$.  The quantity $R$ is called the ``area radius'' and has the property that the area of each symmetry 2-sphere is $4\pi{R^2}$.  Note that in contrast to the Schwarzschild solution we do not choose $R$ as our radial coordinate.  The reason for this is that the numerical evolution method needs the radial coordinate to be spacelike, whereas the area radius $R$ becomes null where an apparent horizon forms.  Therefore numerical simulations that use $R$ as the radial coordinate cannot follow the evolution past the formation of an apparent horizon.  There is still some freedom to choose the radial coordinate.  The numerical method of \cite{dgandted}, which is also our first numerical method, chooses the radial coordinate to be length in the radial direction, which imposes the condition that $\gamma=1$.  Our second numerical method chooses the time evolution vector field to be the aether field $u^a$, which imposes the condition
${\beta^r} =0$.  

Due to spherical symmetry, the extrinsic curvature ${K^a}_b$ has only two independent degrees of freedom, which we choose to be the trace of the extrinsic curvature, denoted $K$, and the radial direction eigenvalue of the trace-free part of the extrinsic curvature, denoted $Q$.  The scalar field degrees of freedom are given by the scalar field $\psi$ and its devivative in the $u^a$ direction, denoted $P$.  It remains to describe the aether field degrees of freedom.  Since we have made the choice that our surfaces of constant time are those orthogonal to $u^a$, most of the information about ${\nabla _a}{u_b}$ is already given by $K_{ab}$.  The only remaining aether field degree of freedom is the acceleration of the aether field ${a_b} \equiv {u^a}{\nabla_a}{u_b}$, which due to spherical symmetry has only a single component $a_r$.

\section{Initial Conditions and Evolution of the Field Equations}

The initial conditions for the scalar field are specified by giving the values at the initial time of $\psi$ and $P$.  Similarly, the initial conditions for the aether field are specified by giving the values at the initial time of $K$ and $a_r$.  We would like those initial conditions to describe a spherical shell that is initially widely enough dispersed so that it is not very strongly gravitating, but also initially radially ingoing so that as it becomes smaller its self gravity may lead to the formation of a black hole.  We choose the following initial data for the scalar field
\bea
\psi  &=& {\frac {a_1} r}(e^{-{(\frac {r - r_{0}}{\sigma})}^2} - e^{-{(\frac {r + {r_0}}{\sigma})}^2}),
\label{psiinit}
\\
P &=& - {\frac {2{a_1}} {\sigma ^2}}\left [ \left ( 1 - {\frac {r_0} r} \right ) e^{-{(\frac {r - r_{0}}{\sigma})}^2} + \left ( 1 + {\frac {r_0} r} \right ) e^{-{(\frac {r + {r_0}}{\sigma})}^2} \right ] ,
\label{Pinit}
\eea 
Here, ${a_1}, \, {r_0}$ and $\sigma$ are constants with the following interpretation: The initial wave forms a spherical shell with a gaussian profile, with $a_1$ the amplitude of the wave, $r_0$ the radius of the shell, and $\sigma$ the width of the shell.  The first term in parentheses in eqn. (\ref{psiinit}) gives the gaussian profile, while the second term is needed to insure that the wave is smooth at $r=0$.  The form of eqn. (\ref{Pinit}) is what is needed to combine with eqn. (\ref{psiinit}) to make the wave purely ingoing.  

We choose the following initial data for the aether field degrees of freedom $a_r$ and $K$
\bea
{a_r}  = - {\frac {2{a_2}} {\sigma ^2}}\left [ \left [ \left ( 1 - {\frac {r_0} r} \right ) + {\frac {\sigma ^2} {2 {r^2}}} \right ]  e^{-{(\frac {r - r_{0}}{\sigma})}^2} - \left [ \left ( 1 + {\frac {r_0} r} \right ) + {\frac {\sigma ^2} {2 {r^2}}} \right ] e^{-{(\frac {r + {r_0}}{\sigma})}^2} \right ] , 
\label{arinit}
\\
K = k {a_2} \left [ \left ( 1 - {\frac {r_0} r} \right ) e^{-{(\frac {r - r_{0}}{\sigma})}^2} + \left ( 1 + {\frac {r_0} r} \right ) e^{-{(\frac {r + {r_0}}{\sigma})}^2} \right ] ,
\label{Kinit}
\eea
Here the constants $r_0$ and $\sigma$ are the same as in eqns. (\ref{psiinit}-\ref{Pinit}), the constant $a_2$ is the amplitude of the aether wave and the constant $k$ is chosen to make the wave purely ingoing.  The reason for this form of the initial data is the following: the equations of motion for $a_r$ and $K$ are similar to those for the wave equation written in first order form with $a_r$ playing the role of the radial derivative of the scalar wave function and $K$ playing the role of the time derivative of the wave function.  Thus if we choose that wave to have the gaussian form of (\ref{psiinit}) with the same radius and width as the scalar wave $\psi$ but with its own independent amplitude, then $a_r$ takes the form of the spatial derivative of that wave and $K$ takes the form that the time derivative of that wave would need to make the wave purely ingoing.

In method 1, we have $\gamma =1$, so in order that both methods compare the same solutions, we choose the initial data for $\gamma$ to be $\gamma =1$ for method 2.  The Hamiltonian constraint equation, eqn. (\ref{ham1}) (which for $\gamma=1$ also provides a solution to eqn. (\ref{ham2})) is then solved numerically to obtain initial data for $R$. 

We now turn to the evolution of the data.  In method 1, the variables $(\psi , P,{a_r},K,R)$ are evolved using eqns. (\ref{dtpsi1},\ref{dtP1},\ref{dtar1},\ref{dtK1},\ref{dtR1}) respectively, while at each time step the variables 
$(Q,\alpha,{\beta^r})$ are obtained by integrating with respect to $r$ eqns. (\ref{drQ1},\ref{dralpha1},\ref{drbetar}) respectively.  In method 2, the variables $(\psi, P, {a_r}, K,R,\gamma)$ are evolved using eqns.  
(\ref{dtpsi2},\ref{dtP2},\ref{dtar2},\ref{dtK2},\ref{dtR2},\ref{dtgamma}) respectively, 
while at each time step the variables 
$(Q,\alpha)$ are obtained by integrating with respect to $r$ eqns. (\ref{drQ2},\ref{dralpha2}) respectively.  

The aether field has a spin-0 mode that travels at speed $v_0$ where 
\cite{spin0}:
\be
{v_0 ^2} = {\frac {{c_{123}}(2-{c_{14}})} {{c_{14}}(1-{c_{13}})(2+{c_{13}}+3{c_2})}} 
\ee
As in the later part of \cite{dgandted} we will choose ${c_3}={c_4}=0$ and will choose $c_2$ so that the speed of the spin-0 mode is 1, which insures that the horizon for the spin-0 mode coincides with the regular horizon.  The condition of unit $v_0$ along with vanishing $c_3$ and $c_4$ yields
\be
{c_2} = {\frac {- {c_1 ^3}} {2-4{c_1} + 3 {c_1 ^2}}}
\label{c2}
\ee 
Thus, all the constants are determined once we pick $c_1$.  

\section{Results}

\begin{figure}
\centering
\includegraphics[width=0.5\textwidth]{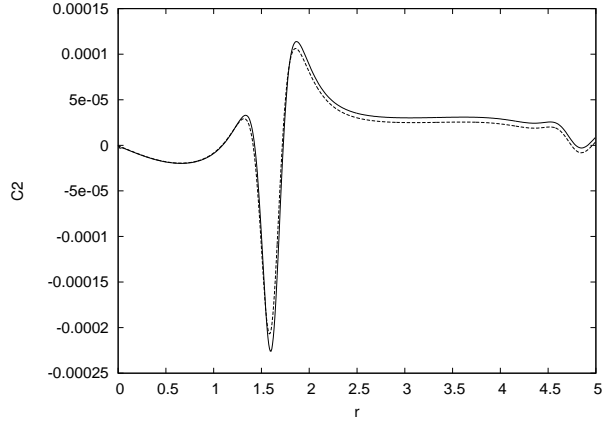}
\caption{for method 1, constraint ${\cal C}_2$ for a simulation with 10,000 points (solid line) and $4{{\cal C}_2}$ for a simulation with 20,000 points (dashed line).  For these simulations we have  ${a_1}=1.5, \, {a_2}=0, \, {r_0}=10, \, \sigma=2, \, {c_1}=0.7$, and outer radius of $80$ and a time of $20$}
\label{cnstrM1}
\end{figure}

\begin{figure}
\centering
\includegraphics[width=0.5\textwidth]{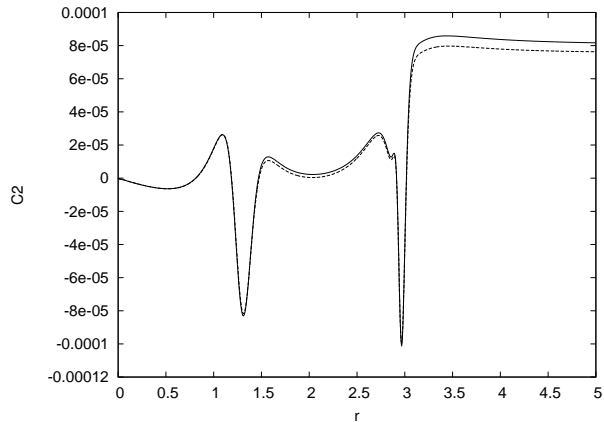}
\caption{for method 2, constraint ${\cal C}_2$ for a simulation with 10,000 points (solid line) and $4{{\cal C}_2}$ for a simulation with 20,000 points (dashed line).  For these simulations we have  ${a_1}=1.5, \, {a_2}=0, \, {r_0}=10, \, \sigma=2, \, {c_1}=0.7$, and outer radius of $80$ and a time of $20$ }
\label{cnstrM2}
\end{figure}

We begin with a convergence test to check the reliability of the simulations.  Recall that there is a constraint quantity ${\cal C}_2$ that vanishes for exact solutions but will not vanish for numerical solutions due to errors involving the finite (space and time) step size.  Because our methods are second order, halving the step size should result in the constraint being four times smaller.  To check this, we run two simulations, one with 10,000 spatial grid points and one with 20,000.  We plot on the same graph, ${\cal C}_2$ for the coarse simulation and $4 {{\cal C}_2}$ for the fine simulation.  Agreement between the two curves indicates second order convergence.  The results are shown in figures \ref{cnstrM1} (for numerical method 1) and \ref{cnstrM2} (for numerical method 2).  The parameters for these simulations are ${a_1}=1.5, \, {a_2}=0, \, {r_0}=10, \, \sigma=2, \, {c_1}=0.7$.  We choose an outer radius of $80$ and run the simulations for a time of $20$.  At this time a weakly gravitating wave would have already dispersed to its original size; but these fields are strongly gravitating and therefore remain confined at smaller radius.  In each case the two curves plotted in the figure agree and thus the code is second order convergent.    

\begin{figure}
\centering
\includegraphics[width=0.5\textwidth]{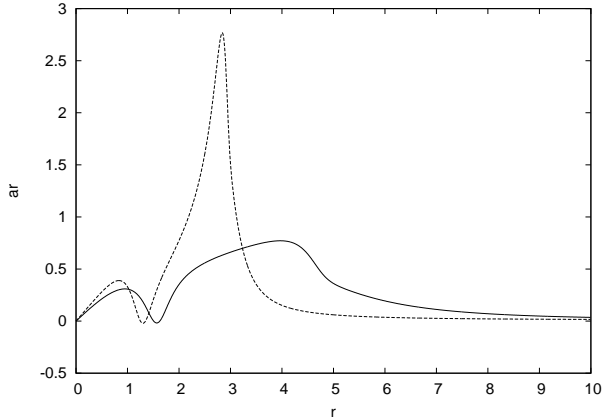}
\caption{a comparison of $a_r$ as a function of $r$ for method 1 (solid line) and method 2 (dashed line).  These simulations are done with 10,000 points, and we have  ${a_1}=1.5, \, {a_2}=0, \, {r_0}=10, \, \sigma=2, \, {c_1}=0.7$, and outer radius of $80$ and a time of $20$ }
\label{acomp1}
\end{figure}

\begin{figure}
\centering
\includegraphics[width=0.5\textwidth]{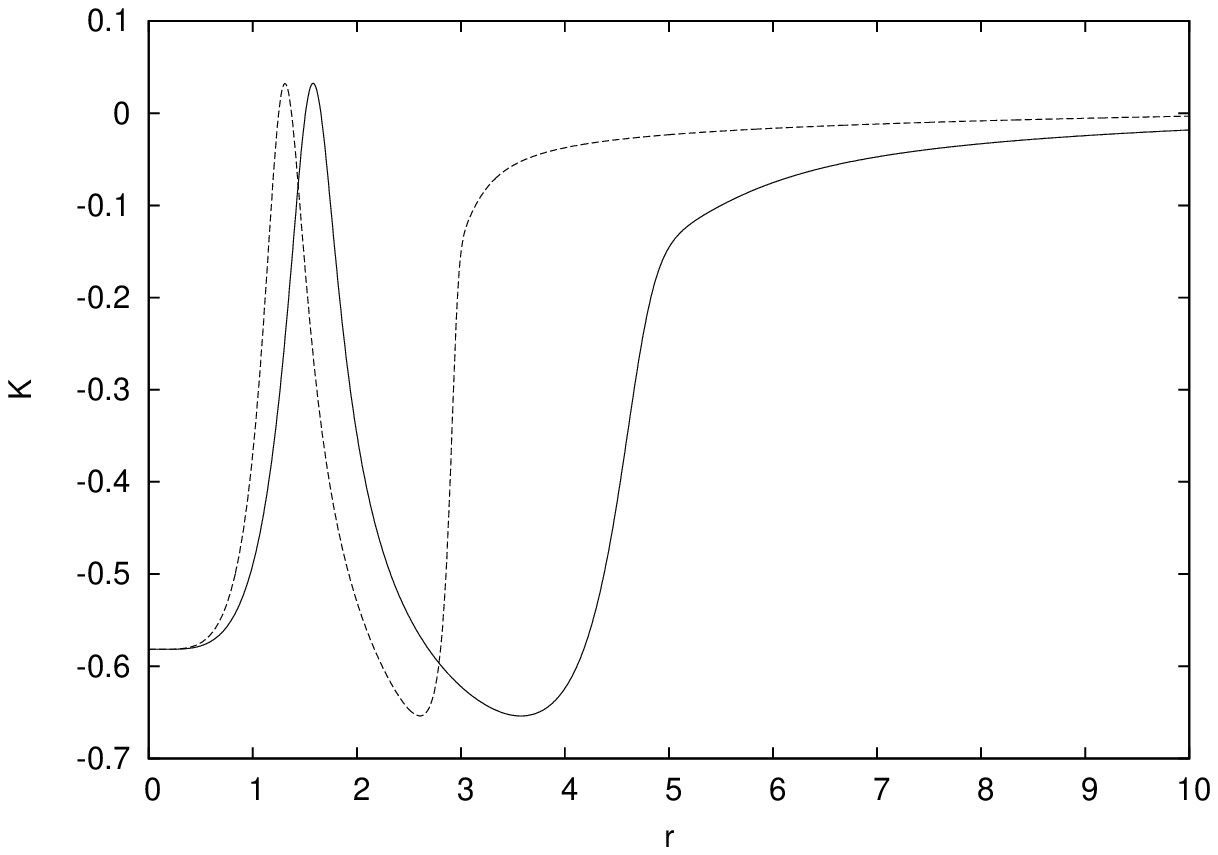}
\caption{a comparison of $K$ as a function of $r$ for method 1 (solid line) and method 2 (dashed line).  These simulations are done with 10,000 points, and we have  ${a_1}=1.5, \, {a_2}=0, \, {r_0}=10, \, \sigma=2, \, {c_1}=0.7$, and outer radius of $80$ and a time of $20$ }
\label{Kcomp1}
\end{figure}

\begin{figure}
\centering
\includegraphics[width=0.5\textwidth]{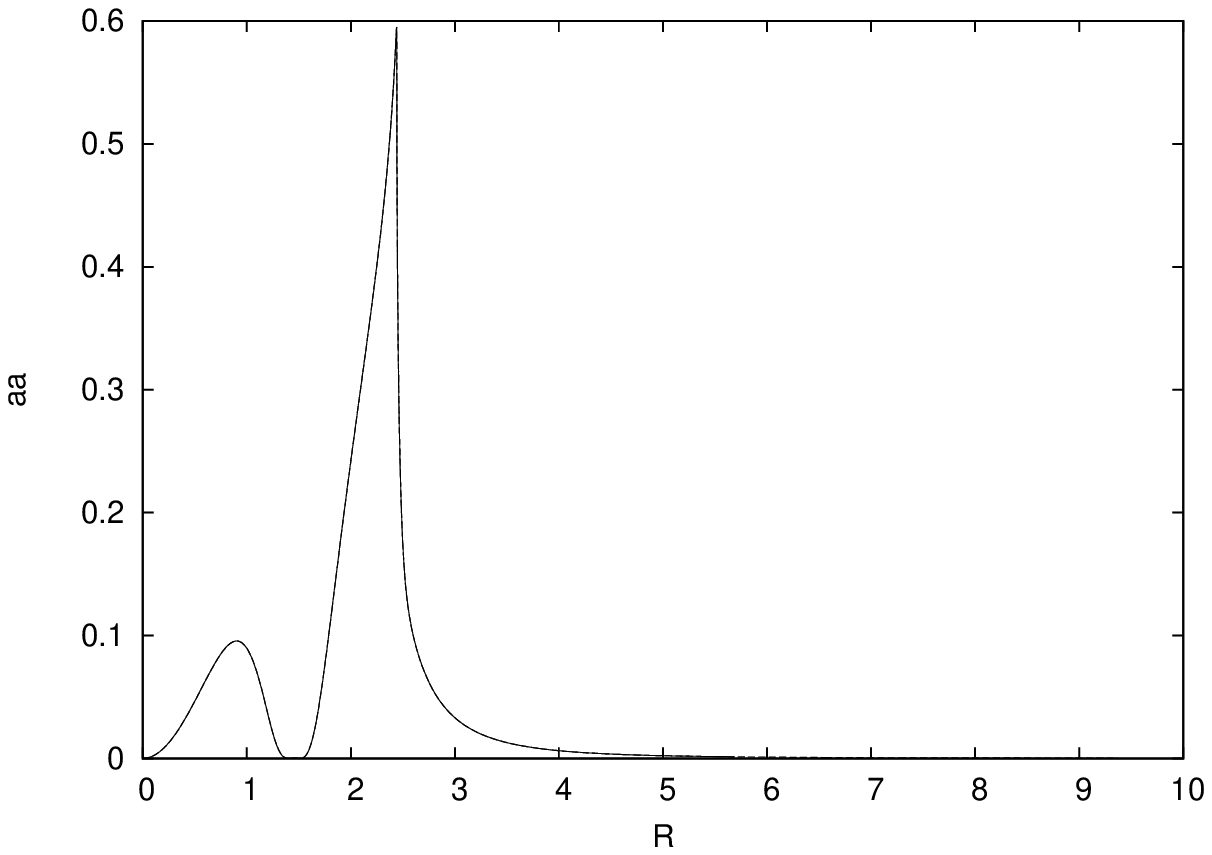}
\caption{a comparison of ${a_a}{a^a}$ as a function of $R$ for method 1 (solid line) and method 2 (dashed line).  These simulations are done with 10,000 points, and we have  ${a_1}=1.5, \, {a_2}=0, \, {r_0}=10, \, \sigma=2, \, {c_1}=0.7$, and outer radius of $80$ and a time of $20$ }
\label{acomp2}
\end{figure}

\begin{figure}
\centering
\includegraphics[width=0.5\textwidth]{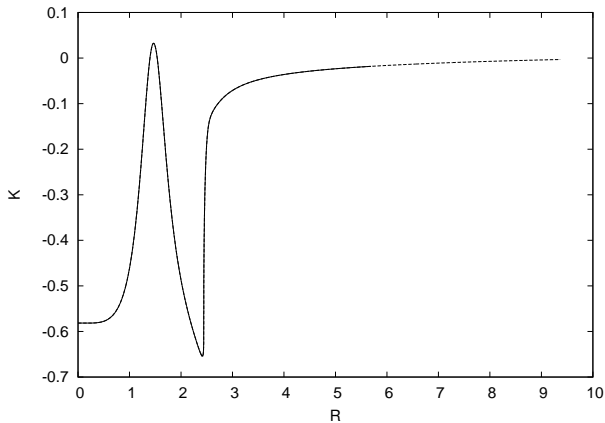}
\caption{a comparison of $K$ as a function of $R$ for method 1 (solid line) and method 2 (dashed line).  These simulations are done with 10,000 points, and we have  ${a_1}=1.5, \, {a_2}=0, \, {r_0}=10, \, \sigma=2, \, {c_1}=0.7$, and outer radius of $80$ and a time of $20$ }
\label{Kcomp2}
\end{figure}

We now consider a comparison between the results of the two methods.  For the same simulation plotted in figs. (\ref{cnstrM1}) and (\ref{cnstrM2}) we plot $a_r$ as a function of $r$ in fig. (\ref{acomp1}) for method 1 and method 2.  The corresponding plot for $K$ as a function of $r$ is given in fig. (\ref{Kcomp1}).  Note that these simulations represent the same situation since they have the same initial data and the same time slicing.  However, the two curves in each figure are different because the $r$ coordinate of method 1 is different from the $r$ coordinate of method 2.  This also means that $a_r$ is different between the two methods, since $a_r$ is the $r$ component of $a_a$.  To make a more direct comparison, we plot invariant quantities.  In particular, note that $K$ (the divergence of the aether field), $R$ (the area radius) and ${a_a}{a^a}$ (the squared magnitude of the acceleration vector) are all invariant quantities.  Figure (\ref{acomp2}) gives a comparison, for the same simulation, of ${a_a}{a^a}$ as a function of $R$ for the two methods.  Note the complete agreement between the two curves.  Figure (\ref{Kcomp2}) performs the same comparison for $K$ as a function of $R$, again with complete agreement.  

\begin{figure}
\centering
\includegraphics[width=0.5\textwidth]{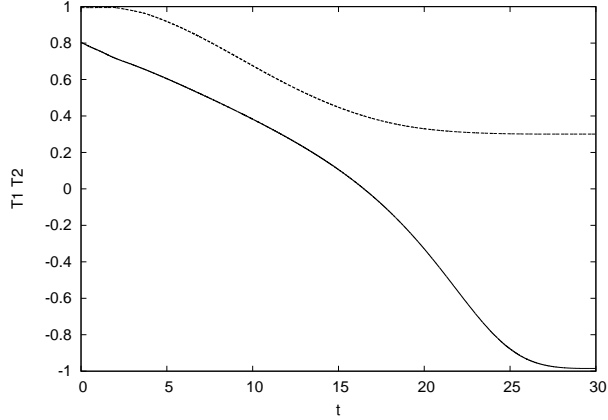}
\caption{minimum values of $T_1$ (solid line) and $T_2$ (dashed line) as a function of time $t$.  A trapped surface forms around 
$t=16.5$. The simulation is done with 10,000 points, and we have  ${a_1}=1.5, \, {a_2}=0, \, {r_0}=10, \, \sigma=2, \, {c_1}=0.7$, and outer radius of $80$. }
\label{bh1}
\end{figure}

\begin{figure}
\centering
\includegraphics[width=0.5\textwidth]{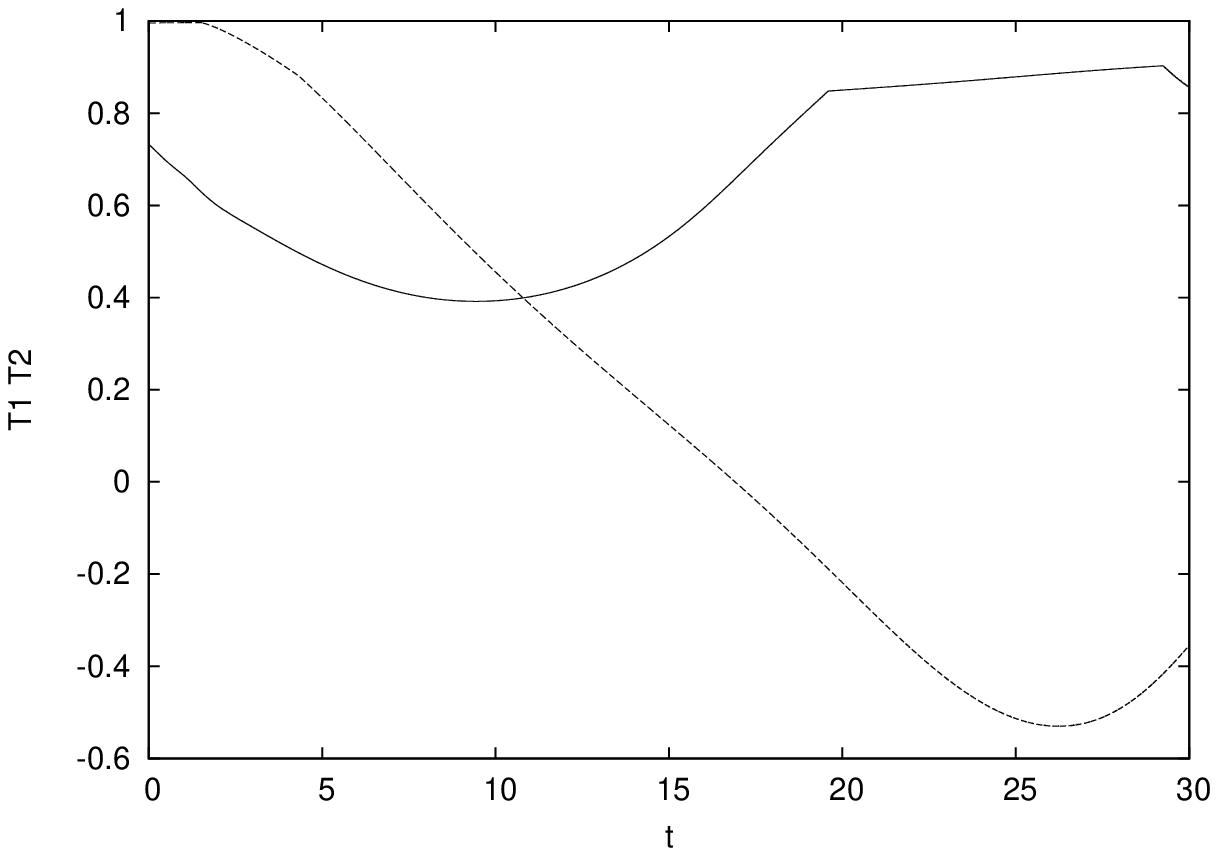}
\caption{minimum values of $T_1$ (solid line) and $T_2$ (dashed line) as a function of time $t$.  An anti-trapped surface forms around 
$t=16.9$. The simulation is done with 10,000 points, and we have  ${a_1}=1.5, \, {a_2}=0, \, {r_0}=10, \, \sigma=2, \, {c_1}=0.8$, and outer radius of $80$. }
\label{wh1}
\end{figure}

\begin{figure}
\centering
\includegraphics[width=0.5\textwidth]{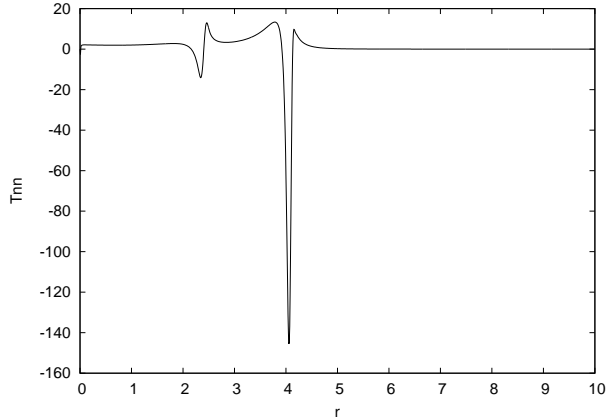}
\caption{Null stress-energy component ${T_{ab}}{n^a}{n^b}$ as a function of $r$ at the time ($t \approx 16.9$) when the anti-trapped surface first forms. The simulation is done using numerical method 2 with 10,000 points, and we have  ${a_1}=1.5, \, {a_2}=0, \, {r_0}=10, \, \sigma=2, \, {c_1}=0.8$, and outer radius of $80$. }
\label{Tnn}
\end{figure}

We now consider the formation of trapped surfaces and anti-trapped surfaces.  In eqn. (\ref{trap12}) we define quantities $T_1$ and $T_2$ that have to do with the expansion of ingoing and outgoing null geodesics.  A trapped surface occurs wherever ${T_1} < 0$ and an anti-trapped surface occurs wherever ${T_2} < 0$.   With the parameters of the previous simulation, we run to a time of $t=30$ and for each time we find the minimum values (over all $r$) of 
$T_1$ and $T_2$.  These results are plotted in fig. (\ref{bh1}).  Note that a trapped forms at $t \approx 16.5$ and remains throughout the rest of the simulation.  No anti-trapped surface forms.  In contrast fig. (\ref{wh1}) shows the results of a simulation with all the parameters of the previous simulation {\emph {except}} that we change $c_1$ from 0.7 to 0.8.  Note that in this case an anti-trapped surface forms at $t \approx 16.9$, while no trapped surface forms.  An anti-trapped surface can form because of violation of the null energy condition.  In particular, what is required is that ${T_{ab}}{n^a}{n^b}$ be negative, where $n^a$ is the ingoing radial null vector.  Figure (\ref{Tnn}) shows ${T_{ab}}{n^a}{n^b}$ at the time at which the anti-trapped surface first forms.   

\begin{figure}
\centering
\includegraphics[width=0.5\textwidth]{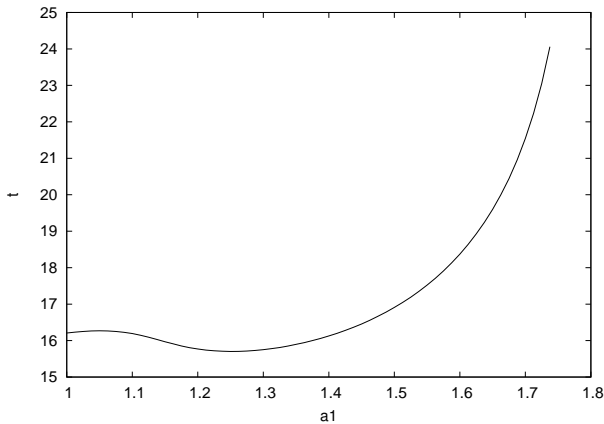}
\caption{Time of formation of an anti-trapped surface as a function of the wave amplitude $a_1$. The simulation is done using numerical method 2 with 10,000 points, and we have  $ {a_2}=0, \, {r_0}=10, \, \sigma=2, \, {c_1}=0.8$, and outer radius of $80$. }
\label{attime1}
\end{figure}

\begin{figure}
\centering
\includegraphics[width=0.5\textwidth]{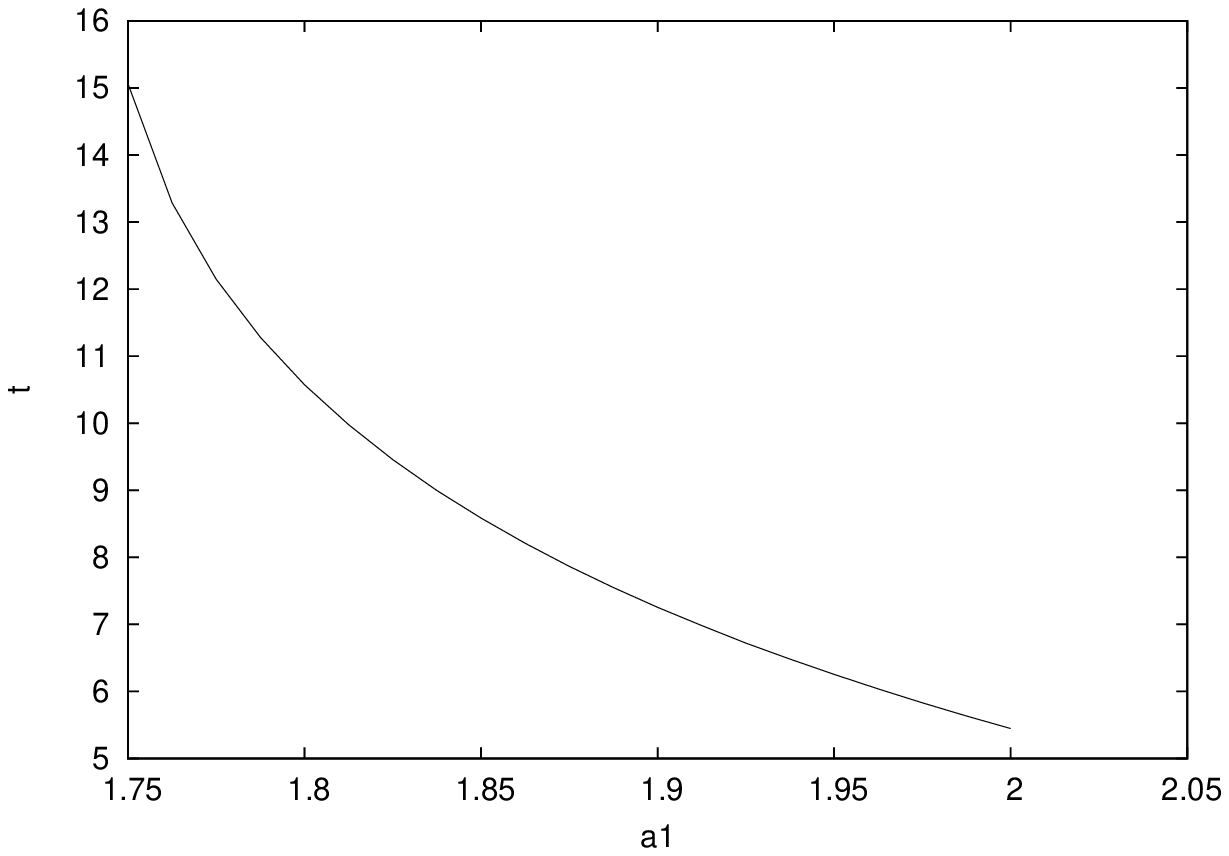}
\caption{Time of formation of a trapped surface as a function of the wave amplitude $a_1$. The simulation is done using numerical method 2 with 10,000 points, and we have  $ {a_2}=0, \, {r_0}=10, \, \sigma=2, \, {c_1}=0.8$, and outer radius of $80$. }
\label{ttime1}
\end{figure}

\begin{figure}
\centering
\includegraphics[width=0.5\textwidth]{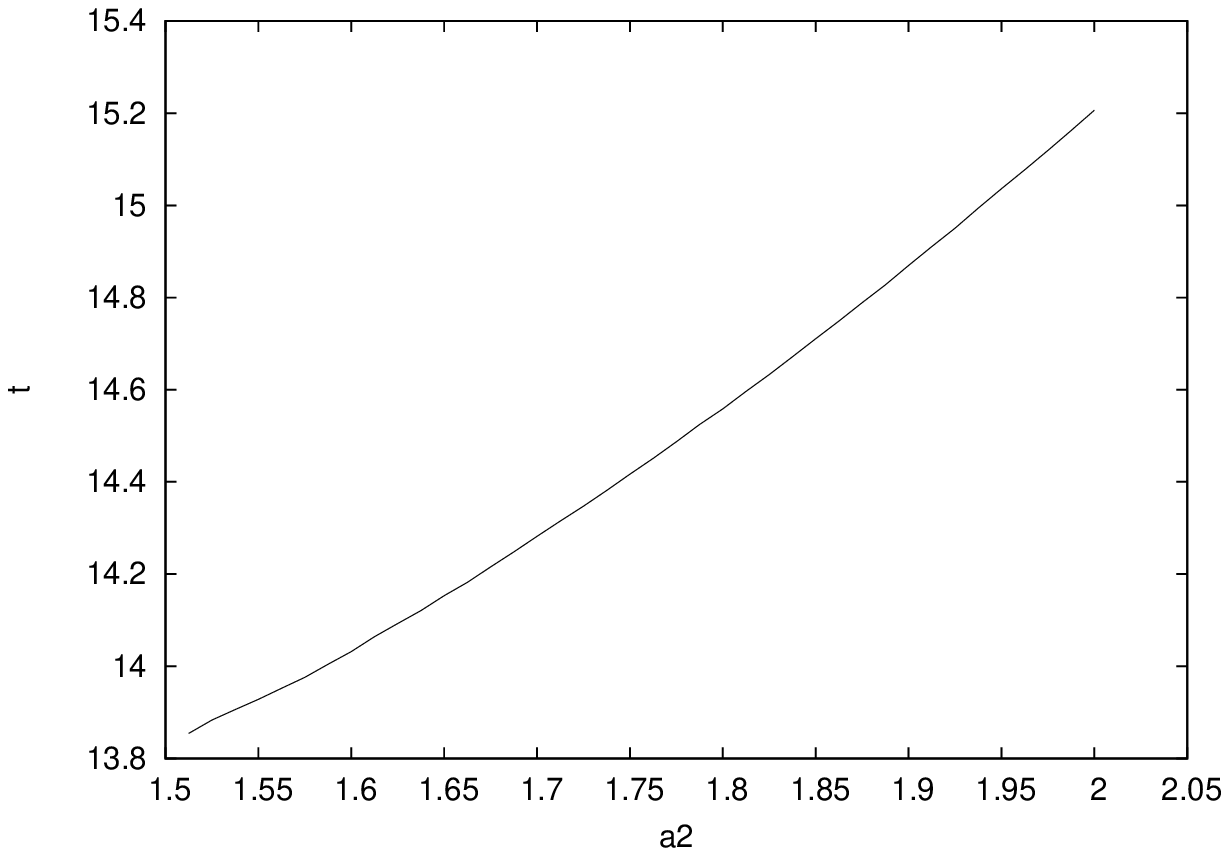}
\caption{Time of formation of an anti-trapped surface as a function of the wave amplitude $a_2$. The simulation is done using numerical method 2 with 10,000 points, and we have  $ {a_1}=0, \, {r_0}=10, \, \sigma=2, \, {c_1}=0.8$, and outer radius of $80$. }
\label{attime2}
\end{figure}

\begin{figure}
\centering
\includegraphics[width=0.5\textwidth]{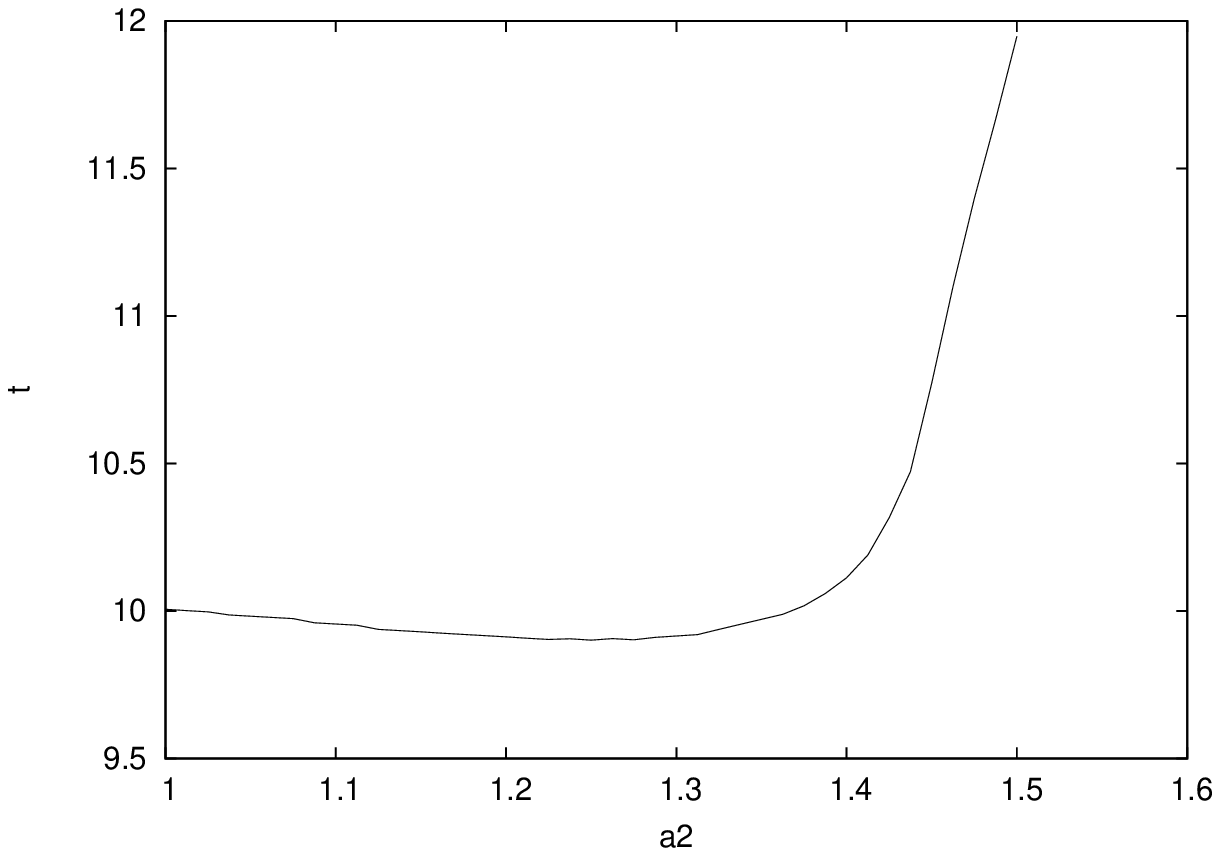}
\caption{Time of formation of a trapped surface as a function of the wave amplitude $a_2$. The simulation is done using numerical method 2 with 10,000 points, and we have  $ {a_1}=0, \, {r_0}=10, \, \sigma=2, \, {c_1}=0.8$, and outer radius of $80$. }
\label{ttime2}
\end{figure}

We now consider how the formation of anti-trapped or trapped surfaces depends on the amplitude of the wave.  We keep all other parameters of the simulation as before, but vary $a_1$ in the range $1 \le {a_1} \le 2$.  For each value of the amplitude, we run the simulation until either a trapped or an anti-trapped surface forms, and then we note the time of formation.  The results of these simulations are shown in figs. (\ref{attime1}) and (\ref{ttime1}).  Here fig. 
(\ref{attime1}) is for those amplitudes for which an anti-trapped surface forms first, and the time of formation of the anti-trapped surface is plotted as a function of the amplitude $a_1$.  Correspondingly, fig. (\ref{ttime1}) graphs the time of formation of a trapped surface as a function of $a_1$ for those values of $a_1$ for which a trapped surface forms first.  In figs. (\ref{attime2}) and (\ref{ttime2}) we display the results of a similar set of simulations where now the scalar wave amplitude $a_1$ is set to zero and the aether wave amplitude $a_2$ is in the range $1 \le {a_2} \le 2$.  In this case, it is the lower amplitude cases in which a trapped surface forms first, and the higher amplitude cases in which an anti-trapped surface forms first.  

\begin{figure}
\centering
\includegraphics[width=0.5\textwidth]{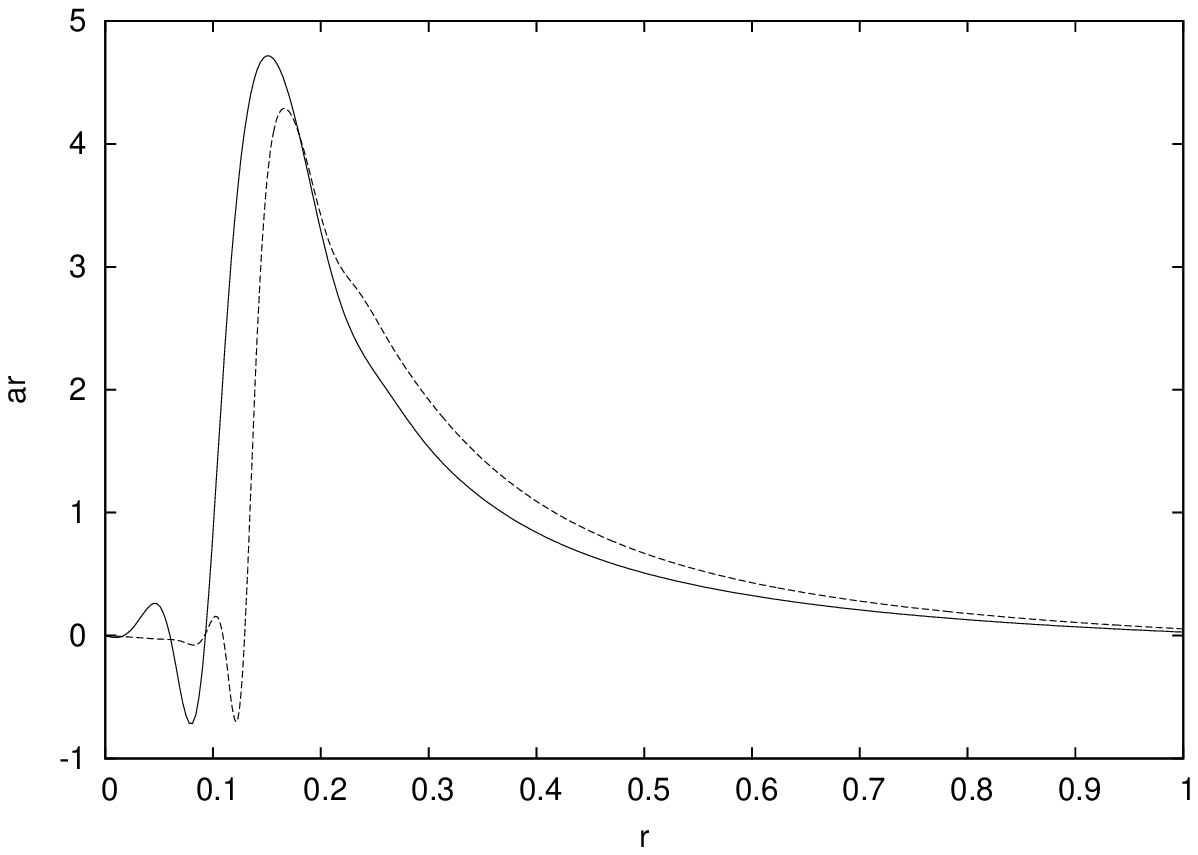}
\caption{$a_r$ as a function of $r$ at two resolutions in the range $0\le r \le 1$ at the time $t \approx 32.3$, which is approximately when $a_r$ and $K$ attain their maximum amplitudes. The simulation is done using numerical method 2 with 40,000 points (solid line) and 80,000 points (dashed line), and we have  ${a_1}=1.5, \, {a_2}=0, \, {r_0}=10, \, \sigma=2, \, {c_1}=0.8$, and outer radius of $120$. }
\label{arres}
\end{figure}

\begin{figure}
\centering
\includegraphics[width=0.5\textwidth]{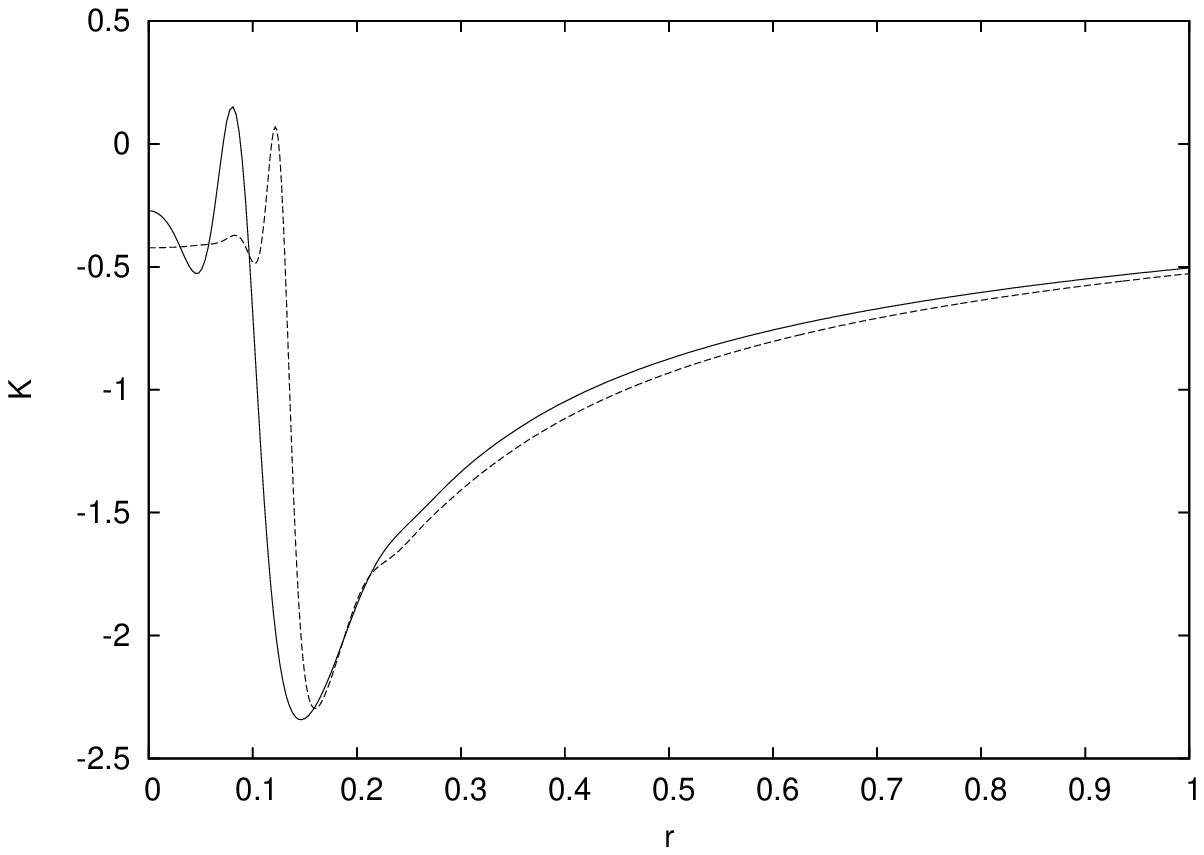}
\caption{$K$ as a function of $r$ at two resolutions in the range $0\le r \le 1$ at the time $t \approx 32.3$, which is approximately when $a_r$ and $K$ attain their maximum amplitudes. The simulation is done using numerical method 2 with 40,000 points (solid line) and 80,000 points (dashed line), and we have  ${a_1}=1.5, \, {a_2}=0, \, {r_0}=10, \, \sigma=2, \, {c_1}=0.8$, and outer radius of $120$. }
\label{Kres}
\end{figure}

We now consider the fate of these white hole spacetimes.  This is somewhat challenging numerically since the simulations must be run for a long enough time for a definitive outcome, and at a high enough resolution to resolve any steep features that develop during the evolution.  In our simulations of the ${c_1}=0.8$ case, we find that eventually both the scalar waves of $\psi$ and $P$ and the aether waves of $K$ and $a_r$ disperse; and the region
in the center settles down to flat spacetime.  However, along the way some sharp features develop.  Figures (\ref{arres}) and (\ref{Kres}) show respectively $a_r$ and $K$, at two different resolutions, for one such simulation at the time when their amplitude is greatest.  These figures indicate that the sharpest features are not fully resolved in this simulation, and therefore conclusions drawn from the simulation may not be completely reliable.  

\section{Conclusions}

Our numerical investigations of gravitational collapse in Einstein-aether theory have turned up an unexpected result: under certain circumstances white holes can form in the collapse process.  However, we note that these white holes differ in several important respects from the textbook white holes of the extended Kruskal diagram of the Schwarzschild spacetime.  In particular, our white holes do not have a singularity in the past, but instead come from the evolution of nonsingular initial data representing a collapsing shell.  This means that to the past of our initial data surface these spacetimes simply consisted of ingoing shells, and the further we go into the past the more highly dispersed and thus less strongly gravitating these shells are.  Our white holes thus manage to evade the Penrose theorem which (in the time reverse of its usually stated form) requires that an anti-trapped surface be accompanied by a singularity to the past.  This evasion is possible because the Penrose theorem posits as a condition that the stress-energy tensor satisfy the null energy condition.  Einstein-aether theory does not satisfy the null energy condition, and therefore the Penrose theorem places no restriction on the formation of white holes in this theory.  We note that some of the theories considered in various cosmological models (such as k-essence or Gallileons) also violate energy conditions.  It would be interesting to see whether white holes can be produced in gravitational collapse in these theories.  

Also, unlike the extended Kruskal diagram, our white holes only last for a finite amount of time.  This is similar to the prediction of Eardley\cite{eardley} for the fate of any white holes in the early universe.  However, both the physical processes and the outcome are somewhat different in our case.  In particular Eardley considers the dynamics of photons in the Kruskal spacetime and conjectures that the white hole will eventually be transformed into a black hole.  In contrast, we treat the dynamics of the aether field and a self-gravitating scalar field, and we find that eventually the fields disperse.

It is somewhat surprising that though \cite{barated} find static black hole solutions even for the case of large $c_1$, our collapse simulations of the large $c_1$ case do not result in the formation of those black holes.  However, we note that static solutions found by solving ODEs can be either stable or unstable; but that the endstate of a collapse process will only result in stable solutions.  This suggests that perhaps the large $c_1$ solutions found in 
\cite{barated} are unstable.  This conjecture could be checked through a numerical treatment of the perturbative modes of those solutions.

Finally, we consider ways of overcoming the numerical challenges of the sharp features that develop during the aether collapse process.  One possibility is to simply perform simulations with a very large number of spatial points.  However, a more efficient solution would be for the simulations to make use of adaptive mesh refinement.

\section*{Acknowledgments}

The work of David Garfinkle is supported in part by NSF grants PHY-1205202 and PHY-1505565.  We would like to thank Ted Jacobson for helpful discussions.

\appendix
\section{Field equations of Einstein-Aether theory}
\label{aeequations}

In this section, we derive the general field equations of Einstein-Aether theory, which we've used to obtain the sets of field equations in the two gauges used.

We use signature $(-+++)$ and the conventions of Wald. The action is

\be
 S = \frac{1}{16\pi G} \int ~d^{4}x  \sqrt{-g}~ \left [ {\cal R}-{K^{ab}{}_{mn}} {{W_a}^m}{{W_b}^n} +
\lambda(g_{ab}u^a u^b + 1) - {\nabla ^a}\psi {\nabla _a}\psi \right ]
\label{action} 
\ee

where ${{W_a}^b}={\nabla _a}{u^b}$, and ${\cal R}$ is the Ricci scalar, $\psi$ is a scalar matter field, $\lambda$ is a Lagrange multiplier to enforce the condition that $u^a$ is a unit vector, and 

\be
{{K^{ab}}_{mn}} = c_1 g^{ab}g_{mn}+c_2\delta^{a}_{m} \delta^{b}_{n}
+c_3\delta^{a}_{n}\delta^{b}_{m}-c_4u^a u^b g_{mn} 
\ee
Here the $c_i$ are dimensionless coupling constants.

Varying the action with respect to the scalar field, we obtain the usual wave equation

\be
{\nabla ^a}{\nabla _a} \psi = 0\label{waveeqn}
\ee

Varying with respect to $u^a$ we obtain the aether equation of motion

\be
{\nabla _a} {{J^a}_b} + \lambda {u_b} + {c_4} {a_a} {\nabla _b} {u^a} = 0
\label{evolveu}
\ee

where ${{J^a}_m} = {{K^{ab}}_{mn}} {{W_b}^n}$ and ${a_a}={u^b}{\nabla _b} {u_a}$.
Varying with respect to the metric, we find the Einstein field equation

\be
{G_{ab}} = {T_{ab}}
\label{EFE}
\ee

where the stress-energy is given by

\bea
{T_{ab}} = - {\textstyle {\frac 1  2}} {g_{ab}}\left ( {{J^c}_d}{{W_c}^d}
+ {\nabla _c}\psi {\nabla ^c} \psi \right )
+ {\nabla _a} \psi {\nabla _b} \psi + {c_4} {a_a}{a_b}
+ \lambda {u_a}{u_b}
\nonumber
\\
 + {c_1} \left ( {W_{ac}}{{W_b}^c} - {W_{ca}}{{W^c}_b} \right ) + {\nabla _c}{F^c _{ab}}
\label{stressEnergy}
\eea

with
\be
{F^c _{ab}} = {{J^c}_{(a}}{u_{b)}} + {u^c}{J_{(ab)}} - {{J_{(a}}^c}{u_{b)}}
\ee

We now specialize to the case where $u^a$ is hypersurface orthogonal (as it always is in spherical symmetry) and use the foliation of spacetime by surfaces orthogonal to $u_a$.  Then the spatial metric $h_{ab}$ and extrinsic curvature $K_{ab}$ are given by

\bea
{h_{ab}} = {g_{ab}}+{u_a}{u_b}
\\
{K_{ab}}= - {{h_a}^c}{\nabla _c}{u_b}
\label{Kdef}
\eea

Then equation (\ref{evolveu}) becomes
\be
{c_{14}} \left ( {{\cal L}_u}{a_a} + 2 {K_{ab}}{a^b} - K {a_a}\right ) + {c_{13}}{D^b}{K_{ab}} + {c_2}{D_a}K = 0
\ee

Here $\cal L$ denotes the Lie derivative, $D_a$ the spatial covariant derivative, and $c_{ik}$ is an abbreviation for ${c_i}+{c_k}$.  

We now use the Einstein field equations to obtain evolution equations and constraint equations for the metric variables. 

The following are standard results of the initial value formulation of general relativity:
\bea
{{\cal L}_u} K = - {D^a}{a_a} - {a^a}{a_a} + {K^{ab}}{K_{ab}} + {\textstyle {\frac 1 2}}(\rho + S)
\label{Ruu}
\\
{D^b}{K_{ab}} - {D_a} K = {j_a}
\label{Gua}
\\
{^{(3)}}{\cal R} + {K^2} - {K^{ab}}{K_{ab}} = 2 \rho
\label{Guu}
\eea

Here ${^{(3)}}{\cal R}$ is the spatial scalar curvature, and the quantities $\rho , \, {j_a}$ and $S$ are given in terms of the stress-energy tensor by  

\bea
\rho \equiv {T_{ab}}{u^a}{u^b}
\\
{j_a} \equiv - {{h_a}^b}{u^c}{T_{bc}}
\\
S \equiv {h^{ab}}{T_{ab}}
\eea

We define $P={u^a}{\nabla _a}\psi$.  Straightforward but tedious algebra using eqns. (\ref{stressEnergy}-\ref{Kdef}) yields the following:

\bea
2 \rho &=& 2 {c_{14}}{D_a}{a^a} + {P^2} + {D_a}\psi {D^a}\psi + {c_{14}}{a_a}{a^a} - {c_2} {K^2} 
- {c_{13}}{K_{ab}}{K^{ab}}
\label{Tuu}
\\
{j_a} &=& - P {D_a}\psi - {c_{14}}(2{K_{ab}}{a^b} - K {a_a} + {{\cal L}_u}{a_a} )
\label{Tua}
\\
2 S &=& 3 {P^2} - {D_a}\psi {D^a}\psi - 2 ({c_{13}}+3{c_2}) {{\cal L}_u} K + {c_{14}}{a_a}{a^a} 
- 3 {c_{13}} {K_{ab}}{K^{ab}} 
\nonumber
\\
&+& (2{c_{13}} + 3 {c_2}){K^2}
\label{Taa}
\eea

Normally, (\ref{Gua}) is considered a constraint equation.  However, in this case $j_a$ has a time derivative.  Nonetheless, using (\ref{Tua}) in (\ref{Gua}) and taking linear combinations with (\ref{evolveu}) we obtain both a constraint equation and an evolution equation:

\bea
(1-{c_{13}}){D^b}{K_{ab}} = (1+{c_2}){D_a}K - P {D_a}\psi
\label{Cmomentum}
\\
{{\cal L}_u}{a_a} = - 2 {K_{ab}}{a^b} + K {a_a} + {{\left [{c_{14}}(1-{c_{13}})\right ] }^{-1}}\left (
{c_{13}} P {D_a}\psi -{c_{123}}{D_a} K \right )
\label{dua}
\eea
Using equations (\ref{Tuu}) and (\ref{Taa}) together in (\ref{Ruu}) we obtain an evolution equation for $K$:

\be
(2+ {c_{13}} + 3 {c_2}) {{\cal L}_u} K = ({c_{14}}-2)({D_a}{a^a} + {a_a}{a^a}) + 2 {P^2} 
+ 2 (1-{c_{13}}){K_{ab}}{K^{ab}} + {c_{123}}{K^2}
\label{duK}
\ee

while using (\ref{Tuu}) in (\ref{Guu}) yields a constraint equation

\be
{^{(3)}}{\cal R} = {c_{14}}(2 {D_a}{a^a} + {a_a}{a^a}) + {P^2} + {D_a}\psi {D^a}\psi + (1-{c_{13}}){K_{ab}}{K^{ab}} 
- (1+{c_2}) {K^2}
\label{CHam}
\ee

We now use the equations of motion to obtain expressions for the stress-energy components that do not involve time derivatives of the fields.  Eqn. (\ref{Tuu}) is already in this form.  Using eqn. (\ref{dua}) in eqn. (\ref{Tua}) we
obtain
\be
{j_a} = {\frac 1 {1-{c_{13}}}} \left ( {c_{123}}{D_a}K - P {D_a}\psi \right )
\label{ja}
\ee
Using eqn. (\ref{duK}) in eqn. (\ref{Ruu}) we obtain
\be
S = - \rho + {\frac 2 {2+{c_{13}}+3{c_2}}} \left [ 2 {P^2} + ({c_{14}}+{c_{13}}+2{c_2})({D_a}{a^a}+{a_a}{a^a}) + {c_{123}}({K^2} - 3 {K_{ab}}{K^{ab}} ) \right ]
\label{S}
\ee

We now consider the properties of trapped and anti-trapped surfaces.  Due to the spherical symmetry, there is a unit vector $s^a$ that is orthogonal to $u^a$ and points in the radial direction.  Then define the null vectors $\ell^a$ and $n^a$ by
\begin{equation}
{\ell ^a} = {u^a} + {s^a} \; , \; \; \; {n^a} = {u^a} - {s^a}
\label{nulldef}
\end{equation}  
That is $\ell ^a$ and $n^a$ are future directed radial null vectors with $\ell ^a$ outgoing and $n^a$ ingoing.  
Now consider the area radius $R$ defined by setting the area of a sphere of symmetry to $4\pi {R^2}$.  Generally, we would expect $R$ to increase along outgoing light rays and decrease along ingoing light rays.  That is, we would expect ${\ell ^a} {\nabla _a} R$ to be positive and ${n^a} {\nabla _a} R$ to be negative.  However under certain circumstances it may happen that ${\ell ^a}{\nabla _a}R$ becomes negative (which is called the presence of a trapped surface) or that ${n^a}{\nabla _a}R$ becomes positive (which is called the presence of an anti-trapped surface). 
In the simulations, we will want to check for the formation of trapped surfaces and anti-trapped surfaces.  We define the quantities $T_1$ and $T_2$ by 
\be
{T_1} = {\ell ^a}{\nabla _a}R \; , \; \; \; {T_2} = - {n^a}{\nabla ^a} R
\label{trap12}
\ee 
Then a trapped surface occurs when ${T_1} < 0 $ and an anti-trapped surface occurs when ${T_2} < 0 $.

What can cause the formation of a trapped surface?  The outgoing light rays can be focussed by stress-energy, more specifically by the component ${T_{ab}}{\ell ^a}{\ell ^b}$ of the stress-energy.  If the outgoing light rays encounter a sufficient amount of positive ${T_{ab}}{\ell ^a}{\ell ^b}$, this can cause the formation of a trapped surface.  Correspondingly, if the ingoing light rays encounter a sufficient amount of {\emph {negative}} 
${T_{ab}}{n^a}{n^b}$, this can defocus them enough to cause the formation of an anti-trapped surface.  Note, however that for any stress-energy satisfying the null energy condition, the quantity ${T_{ab}}{n^a}{n^b}$ cannot be negative.  Furthermore since the null energy condition is implied by the weak, strong, and dominant energy conditions, any stress-energy that satisfies any of those energy conditions also cannot have ${T_{ab}}{n^a}{n^b}$ negative.  Thus, for most physical theories we should not expect the formation of an anti-trapped surface.  (In fact, by the Penrose singularity theorem\cite{penrose} any spacetime satisfying the null energy condition, containing an anti-trapped surface, and having a non-compact Cauchy surface must have a singularity to the {\emph {past}} of the anti-trapped surface).  However, Einstein-aether theory does not satisfy any of these energy conditions, and so the Penrose theorem does not forbid the formation of an anti-trapped surface.  From eqns. (\ref{nulldef}) we obtain the following expression for 
${T_{ab}}{\ell^a}{\ell^b}$ and  ${T_{ab}}{n^a}{n^b}$
\begin{equation}
{T_{ab}}({u^a}\pm {s^a})({u^b}\pm {s^b}) = \rho \mp 2 {j_a}{s^a} + {\textstyle {\frac 1 3}} S + {M_{ab}}{s^a}{s^b}
\label{Tllnn}
\end{equation}  
where the tensor $M_{ab}$ is defined by 
\begin{equation}
{M_{ab}}={{h_a}^c}{{h_b}^d}{T_{ab}} - {\textstyle {\frac 1 3}} S {h_{ab}}
\end{equation}
That is, $M_{ab}$ is the trace-free part of the spatial part of the stress-energy.  From eqn. (\ref{stressEnergy}) we find that $M_{ab}$ can be expressed as
\begin{equation}
{M_{ab}} = {\frac 1 {1-{c_{13}}}} \left [ {L_{ab}} - {\textstyle {\frac 1 3}} L {h_{ab}} \right ]
\label{Mab}
\end{equation}
where the tensor $L_{ab}$ is given by 
\begin{equation}
{L_{ab}} = {D_a}\psi {D_b}\psi + ({c_3}-{c_4}){a_a}{a_b} + {c_{13}} \left ( {D_a}{a_b} - {^{(3)}}{{\cal R}_{ab}} \right )
\label{Lab}
\end{equation}
and ${^{(3)}}{{\cal R}_{ab}}$ is the spatial Ricci tensor.

\section{Field Equations in the first gauge}
\label{gauge1}

Our first numerical method is essentially that of \cite{dgandted}.  We impose spherical symmetry and use as our radial coordinate $r$, which is the length in the radial direction.
Thus the spatial line element takes the form:

\be
d {s^2} = d {r^2} + {R^2} (d {\theta ^2} + {\sin ^2} \theta d {R ^2}).
\ee
Here $R$ is the area radius. The time evolution vector field takes the form 
\be
{t^a} = \alpha {u^a} + {\beta ^a}
\label{g1t}.
\ee
Under spherical symmetry, the only non-zero component of $\beta^a$ is the radial component.

Now, from the definition of P, we have:
\be
{\partial _t}\psi  = \alpha P + {\beta ^r}{\partial _r} \psi
\label{dtpsi1}
\ee 
The wave equation (\ref{waveeqn}) gives us: 
\be
{\partial _t}  P = {\beta ^r}{\partial _r} P + \alpha [PK + {a^r}{\partial _r}\psi  + {\partial _r}{\partial _r}\psi + 2{R^{-1}}{\partial _r}R {\partial _r}\psi ]
\label{dtP1}
\ee

From (\ref{duK}) and (\ref{dua}), we get: 
\bea
{\partial _t} K &=& {\beta ^r}{\partial _r} K + {\frac \alpha 3}{K^2} 
\nonumber
\\
&+& {\frac \alpha {2+{c_{13}}+3{c_2}}}
\left [ ({c_{14}}-2)({\partial _r}{a_r} + 2 {a_r}{R^{-1}}{\partial _r}R + {a_r ^2}) + 2 {P^2} + 3(1-{c_{13}}){Q^2}
\right ]
\label{dtK1}
\eea     
and 
\be
{\partial _t} {a_r} = {\beta ^r} {\partial _r}{a_r} + \alpha \left [ \left ( {\frac {2K} 3} - Q \right ) {a_r} 
+ {\frac {c_{13}} {{c_{14}}(1-{c_{13}})}} P {\partial _r} \psi - {\frac {c_{123}} {{c_{14}}(1-{c_{13}})}} {\partial _r} K \right ]
\label{dtar1}
\ee
respectively. 

Equations (6) and (\ref{g1t}) together give: 
\be
{{\cal L}_t} h_{ab} = -2\alpha K_{ab} + {{\cal L}_\beta} h_{ab}.
\ee
The $\theta \theta$ and $rr$ components of  this equation are respectively: 
\bea
{\partial _t}R = {\beta ^r}{\partial _r}R + \alpha R \left ( {\frac Q 2} - {\frac K 3} \right )
\label{dtR1}
\\
{\partial _r}{\beta ^r} = \alpha \left ( Q + {\frac K 3} \right )
\label{drbetar}
\eea
where Q = ${{K^r}_r}-K/3$ is the trace-free part of the extrinsic curvature. 

Equation (\ref{Cmomentum}) gives us:
\be
{\partial _r} Q = - 3 Q {R^{-1}} {\partial _r}R + {{(1-{c_{13}})}^{-1}} \left [ {\textstyle {\frac 1 3}} (2 + {c_{13}} + 3 {c_2}) {\partial _r} K - P {\partial _r} \psi \right ]
\label{drQ1}
\ee

Now, from the definition of $\alpha$, we have: 
\be
{\partial _r} \ln \alpha = {a_r}
\label{dralpha1}
\ee

The Hamiltonian initial value constraint given by (\ref{CHam}) then becomes the vanishing of the quantity $\cal C$ given by 
\bea
{\cal C} = {\partial _r}{\partial _r} R + {\frac {{{({\partial _r}R)}^2} - 1} {2R}} + {c_{14}} {a_r}{\partial _r}R
\nonumber
\\
+ {\frac R 4} \left [ {c_{14}}(2{\partial _r}{a_r}+{a_r}{a_r}) + {P^2} + {{({\partial _r}\psi )}^2} +
{\textstyle {\frac 3 2}}(1-{c_{13}}){Q^2} - {\textstyle {\frac 1 3}}(2+{c_{13}}+3{c_2}){K^2} \right ]
\label{ham1}
\eea
It will be helpful to use $\cal C$ to define a related constraint quantity ${\cal C}_2$ given by
\be
{{\cal C}_2} = - 2 {\int _0 ^r} {\cal C} R {\partial _r} R \; dr 
\ee
Then using eqn. (\ref{ham1}) we obtain
\bea
{{\cal C}_2} = R (1-{{({\partial _r}R)}^2}) - {\int _0 ^r} d r \; {\partial _r} R \; \biggl [ 2 {c_{14}} {a_r} R {\partial _r}R 
\nonumber
\\ 
+  {\frac {R^2} 2} \left [ {c_{14}}(2{\partial _r}{a_r}+{a_r}{a_r}) + {P^2} + {{({\partial _r}\psi )}^2} +
{\textstyle {\frac 3 2}}(1-{c_{13}}){Q^2} - {\textstyle {\frac 1 3}}(2+{c_{13}}+3{c_2}){K^2} \right ]
\biggr ]
\label{hamc21}
\eea

We now consider the formation of trapped surfaces and anti-trapped surfaces.  The unit radial vector $s^a$ has component ${s^r} =1$.  Then using eqn. (\ref{dtR1}) in eqn. (\ref{trap12}) we obtain
\bea
{T_1} = {\partial _r} R \; + \; R \left ( {\frac Q 2 } - {\frac K 3} \right )
\label{T11}
\\
{T_2}= {\partial _r} R \; - \; R \left ( {\frac Q 2 } - {\frac K 3} \right )
\label{T21}
\eea   

We now assemble expressions for the stress-energy components needed to calculate ${T_{ab}}{\ell^a}{\ell^b}$ and
${T_{ab}}{n^a}{n^b}$ via eqn. (\ref{Tllnn}).  From eqns. (\ref{Tuu},\ref{ja},\ref{S}) we obtain
\bea
2 \rho &=&  {c_{14}} \left (  2{\partial _r}{a_r} + {a_r} [ 4 {R^{-1}}{\partial _r}R +{a_r} ]   \right ) + {P^2} + {{({\partial _r}\psi )}^2} 
\nonumber
\\
&-& {\textstyle {\frac 1 3}}( {c_{13}}+3{c_2}) {K^2} 
- {\textstyle {\frac 3 2}}{c_{13}}{Q^2}
\label{Tuu1}
\\
{j_a}{s^a} &=& {\frac 1 {1-{c_{13}}}} \left ( {c_{123}}{\partial _r}K - P {\partial _r}\psi \right )
\label{ja1}
\\
S &=& - \rho + {\frac 2 {2+{c_{13}}+3{c_2}}} \biggl [ 2 {P^2} - {\textstyle {\frac 9 2}} {c_{123}}{Q^2}
\nonumber
\\
&+& ({c_{14}}+{c_{13}}+2{c_2})  
\left (  {\partial _r}{a_r} + {a_r} [ 2 {R^{-1}}{\partial _r}R +{a_r} ]   \right )  \biggr ]
\label{S1}
\eea
Using eqns. (\ref{Mab}) and (\ref{Lab}) we obtain
\bea
{M_{ab}}{s^a}{s^b} &=& {\frac 2 {3(1-{c_{13}})}} \biggl [ {{({\partial _r}\psi)}^2} + ({c_3}-{c_4}){{({a_r})}^2}
\nonumber
\\
&+& {c_{13}} \left ( {\partial _r}{a_r} +{a_r}{R^{-1}}{\partial _r}R + {R^{-1}}{\partial _r}{\partial _r}R 
+ {R^{-2}}[1- {{({\partial _r}R)}^2}] \right ) \biggr ]
\label{M1}
\eea

\section{Field Equations in the second gauge}
\label{gauge2}

Our second numerical method is to impose spherical symmetry and use zero shift.  That is, evolution takes place in the direction of the aether field.  The spaceime line element then takes the form
\be
d {s^2} = - {\alpha ^2} d {t^2} + \gamma d {r^2} + {R^2} (d {\theta ^2} + {\sin ^2} \theta d {\phi ^2} ) 
\ee
In terms of components of $h_{ab}$ we have 
$ {h_{rr}}=\gamma, \, {h_{\theta \theta}}= {R^2}$ and ${h_{\phi \phi}} = {R^2} {\sin ^2} \theta$.  
Thus, in comparison to the previous method we no longer have a shift, but now the radial component of the metric is a degree of freedom.
The time evolution vector field takes the form ${t^a} = \alpha {u^a}$.  From the definition of $P$ we find
\be
{\partial _t} \psi = \alpha P 
\label{dtpsi2}
\ee
while from the wave equation we obtain
\be
{\partial _t} P = \alpha P K + {\frac \alpha \gamma} \left [ {\partial _r} {\partial _r}\psi 
+ \left ( {a^r} + 2 {R^{-1}} {\partial _r} R - {\textstyle {\frac 1 2}} {\gamma ^{-1}} {\partial _r}\gamma 
\right ) {\partial _r}\psi \right ]
\label{dtP2}
\ee
It is helpful to define $A_{ab}$ to be the trace-free part of $K_{ab}$ and to define $Q \equiv {{A^r}_r}$.  From the definition of extrinsic curvature and the vanishing of the shift we have
\be
{{\cal L}_t} {h_{ab}} = - 2 \alpha {K_{ab}} 
\label{dth2}
\ee
The $rr$ component of eqn. (\ref{dth2}) yields
\be
{\partial _t}\gamma = - 2 \alpha \gamma (Q + K/3)
\label{dtgamma}
\ee
while the $\theta \theta$ component yields
\be
{\partial _t}R = \alpha R (Q/2 - K/3)
\label{dtR2}
\ee
From the definition of $\alpha$ we have
\be
{\partial _r} \ln \alpha = {a_r}
\label{dralpha2}
\ee
Equation (\ref{Cmomentum}) yields 
\be
{\partial _r} Q = - {\frac {3 Q} R} {\partial _r} R + {{(1-{c_{13}})}^{-1}} \left [ {\textstyle {\frac 1 3}}
(2 + {c_{13}} + 3 {c_2}){\partial _r} K - P {\partial _r} \psi \right ]
\label{drQ2}
\ee
while eqns. (\ref{dua}) and (\ref{duK}) yield respectively 
\bea
{\partial _t}{a_r} &=& \alpha \left [ (K/3 - 2Q) {a_r} + {{\left [ {c_{14}}(1-{c_{13}}) \right ] }^{-1}} \left ( {c_{13}} P {\partial _r} \psi - {c_{123}} {\partial _r} K \right ) \right ]
\label{dtar2}
\\
{\partial _t} K &=&  {\frac \alpha 3}{K^2}
+ {\frac \alpha {2+{c_{13}} + 3 {c_2}}} \biggl [ 2 {P^2} + 3 (1-{c_{13}}){Q^2}
\nonumber
\\
&+& {\frac {{c_{14}}-2} \gamma} \left ( {\partial _r}{a_r} + (2 {R^{-1}}{\partial _r}R - {\textstyle {\frac 1 2}}
{\gamma ^{-1}} {\partial _r}\gamma + {a_r}) {a_r} \right )  \biggr ] 
\label{dtK2}
\eea
The Hamiltonian constraint (eqn. (\ref{CHam})) then becomes the vanishing of the quantity $\cal C$ given by
\bea
{\cal C} &=& {\partial _r}{\partial _r} R + {\frac {{{({\partial _r}R )}^2} - \gamma} {2 R}} - {\frac 1 {2\gamma}}
{\partial _r}\gamma {\partial _r} R
\nonumber
\\
&+& {\frac R 4} \left [  {c_{14}} \left  (2 {\partial _r}{a_r} + (4 {R^{-1}}{\partial _r}R - 
{\gamma ^{-1}}{\partial _r} \gamma +{a_r}){a_r}) \right ) + {{({\partial _r}\psi )}^2}  \right ]
\nonumber
\\
&+& {\frac {R\gamma} 4} \left [  {P^2} + {\textstyle {\frac 3 2}} (1-{c_{13}}){Q^2} - {\textstyle {\frac 1 3 }}
(2 + {c_{13}} + 3 {c_2}){K^2} \right ]
\label{ham2}
\eea
It will be helpful to use $\cal C$ to define a related constraint quantity ${\cal C}_2$ given by
\be
{{\cal C}_2} = - 2 {\int _0 ^r} {\cal C} {\gamma ^{-1}} R {\partial _r} R \; dr 
\ee
Then using eqn. (\ref{ham2}) we obtain
\bea
{{\cal C}_2} &=& R (1-{\gamma ^{-1}} {{({\partial _r}R)}^2}) 
\nonumber
\\
&-& {\int _0 ^r} dr \, ({\partial _r}R) \, \biggl [ 
{\frac {R^2} {2\gamma}} \left [  {c_{14}} \left  (2 {\partial _r}{a_r} + (4 {R^{-1}}{\partial _r}R - 
{\gamma ^{-1}}{\partial _r} \gamma +{a_r}){a_r}) \right ) + {{({\partial _r}\psi )}^2}  \right ]
\nonumber
\\
&+& {\frac {R^2} 2} \left [  {P^2} + {\textstyle {\frac 3 2}} (1-{c_{13}}){Q^2} - {\textstyle {\frac 1 3 }}
(2 + {c_{13}} + 3 {c_2}){K^2} \right ] \biggr ]
\label{hamc22}
\eea

We now consider the formation of trapped surfaces and anti-trapped surfaces.  The unit radial vector $s^a$ has component ${s^r} ={\gamma ^{-1/2}}$.  Then using eqn. (\ref{dtR2}) in eqn. (\ref{trap12}) we obtain
\bea
{T_1} = {\partial _r} R \; + \; R \left ( {\frac Q 2 } - {\frac K 3} \right )
\label{T12}
\\
{T_2}= {\partial _r} R \; - \; R \left ( {\frac Q 2 } - {\frac K 3} \right )
\label{T22}
\eea   

We now assemble expressions for the stress-energy components needed to calculate ${T_{ab}}{\ell^a}{\ell^b}$ and
${T_{ab}}{n^a}{n^b}$ via eqn. (\ref{Tllnn}).  From eqns. (\ref{Tuu},\ref{ja},\ref{S}) we obtain
\bea
2 \rho &=&  {c_{14}} {\gamma ^{-1}} \left (  2{\partial _r}{a_r} + {a_r} [ 4 {R^{-1}}{\partial _r}R 
-{\gamma ^{-1}} {\partial _r}\gamma +{a_r} ]   \right ) + {P^2} + {\gamma ^{-1}} {{({\partial _r}\psi )}^2} 
\nonumber
\\
&-& {\textstyle {\frac 1 3}}( {c_{13}}+3{c_2}) {K^2} 
- {\textstyle {\frac 3 2}}{c_{13}}{Q^2}
\label{Tuu2}
\\
{j_a}{s^a} &=& {\frac {\gamma ^{-1/2}} {1-{c_{13}}}} \left ( {c_{123}}{\partial _r}K - P {\partial _r}\psi \right )
\label{ja2}
\\
S &=& - \rho + {\frac 2 {2+{c_{13}}+3{c_2}}} \biggl [ 2 {P^2} - {\textstyle {\frac 9 2}} {c_{123}}{Q^2}
\nonumber
\\
&+& ({c_{14}}+{c_{13}}+2{c_2}) {\gamma ^{-1}} 
\left (  {\partial _r}{a_r} + {a_r} [ 2 {R^{-1}}{\partial _r}R - {\textstyle {\frac 1 2}}{\gamma ^{-1}}{\partial _r}\gamma +{a_r} ]   \right )  \biggr ]
\label{S2}
\eea
Using eqns. (\ref{Mab}) and (\ref{Lab}) we obtain
\bea
{M_{ab}}{s^a}{s^b} &=& {\frac {2 {\gamma ^{-1}}} {3(1-{c_{13}})}} \biggl [ {{({\partial _r}\psi)}^2} + ({c_3}-{c_4}){{({a_r})}^2}
+ {c_{13}} \bigl ( {\partial _r}{a_r} +{a_r}({R^{-1}}{\partial _r}R - {\textstyle {\frac 1 2}} {\gamma ^{-1}}{\partial _r}\gamma ) 
\nonumber
\\
&+& {R^{-1}}{\partial _r}{\partial _r}R - {\textstyle {\frac 1 2}} {\gamma ^{-1}}{R^{-1}}{\partial _r}\gamma {\partial _r}R
+ {R^{-2}}[\gamma - {{({\partial _r}R)}^2}] \bigr ) \biggr ]
\label{M2}
\eea

\end{document}